\documentclass[letterpaper, 10 pt, conference]{ieeeconf}  

\IEEEoverridecommandlockouts                              

\usepackage{cite}
\usepackage{amsmath,amssymb,amsfonts}
\usepackage{algorithmic}
\usepackage{graphicx}
\usepackage{textcomp}
\usepackage{subcaption} 
\usepackage{xcolor}
\def\BibTeX{{\rm B\kern-.05em{\sc i\kern-.025em b}\kern-.08em
    T\kern-.1667em\lower.7ex\hbox{E}\kern-.125emX}}

\usepackage{orcidlink}
\usepackage{iftex} 

\title{\LARGE \bf
Physiological Measures of the Mental Workload in Users of a Lower Limb Exosuit: A Comparison of Subjective and Objective Metrics}

\author{Giulia Mariani*$^{1,2}$, Chiara Lambranzi*$^{3,4}$, Nicholas Cartocci*$^{2,3}$,\\ Giacinto Barresi$^{1,5}$, Christian Di Natali$^{3}$, Elena De Momi$^{4}$ and Jesus Ortiz$^{3}$ 
\thanks{*These authors contributed equally to this work.}%
\thanks{$^{1}$Rehab Technologies Lab, Istituto Italiano di Tecnologia, Genoa, Italy.}%
\thanks{$^{2}$Department of Informatics, Bioengineering, Robotics, and Systems Engineering, University of Genoa, Genoa, Italy.}%
\thanks{$^{3}$Advanced Robotics, Istituto Italiano di Tecnologia, Genoa, Italy.}%
\thanks{$^{4}$Department of Electronics, Information and Bioengineering, Politecnico di Milano, Milan, Italy.}%
\thanks{$^{5}$Bristol Robotics Laboratory, University of the West of England, Bristol, United Kingdom.}%
}

\begin{document}

\maketitle
\thispagestyle{empty}
\pagestyle{empty}

\begin{abstract}

Lower-limb exosuits are particularly relevant for individuals with some degree of mobility impairment, such as post-stroke patients or older adults with reduced movement capabilities. This study aims to investigate the mental workload (MWL) assessment of XoSoft, a lower-limb soft exoskeleton, using and comparing subjective and objective physiological metrics. The NASA-TLX questionnaire, the average percentage change in pupil size (APCPS), and the Baevsky stress index (SI) are compared. The experiments were conducted on 18 healthy subjects while walking and involved mathematical tasks to create a double-task condition. The results show a complex interaction between task difficulty, exoskeleton activation, and pupillary dynamics, suggesting that the subject might reach a saturated condition under a high mental load. Besides, the data indicate that pupil diameter may be an objective mental workload indicator that correlates with subjective NASA-TLX questionnaires.
The discordant indications from the stress index suggest how different metrics of the ocular and cardiac levels respond differently to various stimuli and dynamics. Research has also revealed ocular asymmetry, with the right eye more sensitive to cognitive load.

\end{abstract}


\section{Introduction}

Mental workload (MWL) can be defined as "the portion of operator information processing capacity or resources that is actually required to meet system demands" \cite{Eggemeier2020} or also as "the cost of performing a task in terms of a reduction in the capacity to perform additional tasks that use the same processing resource" \cite{Kramer1987}. These definitions merge the classical-psychological view that connects MWL to the concept of attention and performance and the scientific-engineering perspective that connects it to a concept of effectiveness of the system \cite{Huey}.
It is a wide concept, studied and analyzed in different contexts, that is not easy to identify and does not have a universal definition \cite{Young_2015, Cain_2007}. MWL depends on many parameters, including the difficulty level of the task (requests, performances, etc.), the subjective user experience (skill, attention, etc.), the execution time, the overload, the environmental context in which performance occurs, stress, fatigue, and all these factors make measurement difficult \cite{Cain_2007}.

The study of MWL is essential in understanding human performance, making it one of the most significant topics in psychology, ergonomics, and human factors. It is therefore a critical consideration in the design and evaluation of human-machine systems. For instance, depending on the user’s experience and the complexity of the task, the MWL associated with using an assistance device may lead to performance inhibition, potentially resulting in discarding the device altogether \cite{Batavia_1990}.

One domain where MWL plays a crucial role is in the development of wearable assistive technologies, such as exoskeletons and exosuits. 
For the populations that benefit from clinical lower-limb exoskeletons, the cognitive effort required to operate an assistive device is a critical factor. Walking is already a cognitively demanding task, especially in older adults, who experience age-related declines in cognitive processing, balance control, and motor coordination \cite{Lindenberger_2000}. When MWL is too high, it can negatively affect balance, leading to increased postural sway, delayed reactions to external perturbations, and an overall reduction in stability—factors that heighten the risk of falls, particularly when mental fatigue sets in \cite{Pitts_2022}. This underscores the need for assistive devices that minimize MWL, ensuring accessibility, safety, and ease of use for end users.

Measuring MWL in the context of wearable assistive devices is a challenge, and various assessment techniques have been developed. These can be broadly categorized into four approaches, according to the review by Marchand et al.\cite{marchand_2021}: subjective assessments, secondary task procedures, physiological measures, and modeling.
Subjective assessments are based on questionnaires in which the participant assesses their own mental workload through rating scales such as the NASA Task Load Index (NASA-TLX) or the System Usability Scale (SUS). While these tools are widely used, they are inherently subjective and can be influenced by personal biases, mood, and inconsistent interpretations of rating scales.
In the double-task paradigm, the participant has to perform two tasks simultaneously, and the performance on the secondary one reflects the mental workload induced by the primary task. However, this method lacks standardization across studies, making it difficult to compare results \cite{EsmaeiliBijarsari2021}. While these two types of assessment are the most widely used in works that evaluate wearable assistive devices, other approaches may overcome some of the limitations that they present.
Physiological measures are objective and can be standardized and compared in different studies, but there is no objective metric that can be considered the gold standard. There are several physiological measures that can be used, such as electrocardiographic measurements (ECG), electroencephalographic measurements (EEG), electromyographic measurements (EMG), breathing measures, and measures of eye movement. 

This study aims to investigate the impact of LLE design and use on the MWL of healthy individuals, with a specific focus on objective physiological markers. While subjective assessments remain common, they are often prone to variability and potential bias. Similar to the research in \cite{Ahmadi_2024}, we seek to compare these traditional methods with two physiological metrics that offer an objective measure of MWL:
\begin{itemize}
    \item Pupillometry: Pupil dilation correlates with cognitive effort, as the autonomic nervous system regulates pupil size in response to mental workload. An increase in pupil dilation typically indicates a higher cognitive load.
    \item Heart Rate (HR): HR reflects the balance between the sympathetic and parasympathetic nervous systems and is commonly used as an indicator of stress and cognitive demand. An increase in stress is often observed with a higher mental workload.
\end{itemize}

The goal is to assess how these physiological markers (pupillometry and HR) perform in assessing MWL and how they correlate with subjective reports in order to find objective, reliable methods to evaluate MWL in settings involving LLE.
This study contributes by being the first to jointly compare subjective MWL reports with both pupillometry and heart rate metrics in the context of LLE. Furthermore, it is among the few studies in the exoskeleton field to explore MWL using ocular metrics, offering new and relevant insights for researchers aiming to adopt these measures in assisted locomotion tasks.
Given the primary objective of this study, the performance data of the double task experiments are not presented and will be addressed in future work.

\section{Materials and Methods}
This section describes the experimental setup, with the exosuit and the sensors, and the procedure used in the study.
The exosuit used is XoSoft Gamma, developed during the XoSoft EU Project. In the experiments, the exosuit was tested in two ways: inactive, the device is worn but does not provide assistance, and active. In this way, it was possible to evaluate the additional mental workload resulting from the device’s weight and bulk, that are a constant characteristic of XoSoft, and noise of the pneumatic system, that is present only when the exosuit is active. The experiment analyzes questionnaires and physiological signals: pupillometry and heart rate. Pupillometry data are collected using the Tobii Pro Glasses 3 eye tracker (Tobii AB, Stockholm, Sweden), while HR was gathered from the Empatica E4 wristband sensor (Empatica Inc., Cambridge, MA, USA).

\begin{figure*}[ht!]
    \centering \centerline{\includegraphics[width=\linewidth]{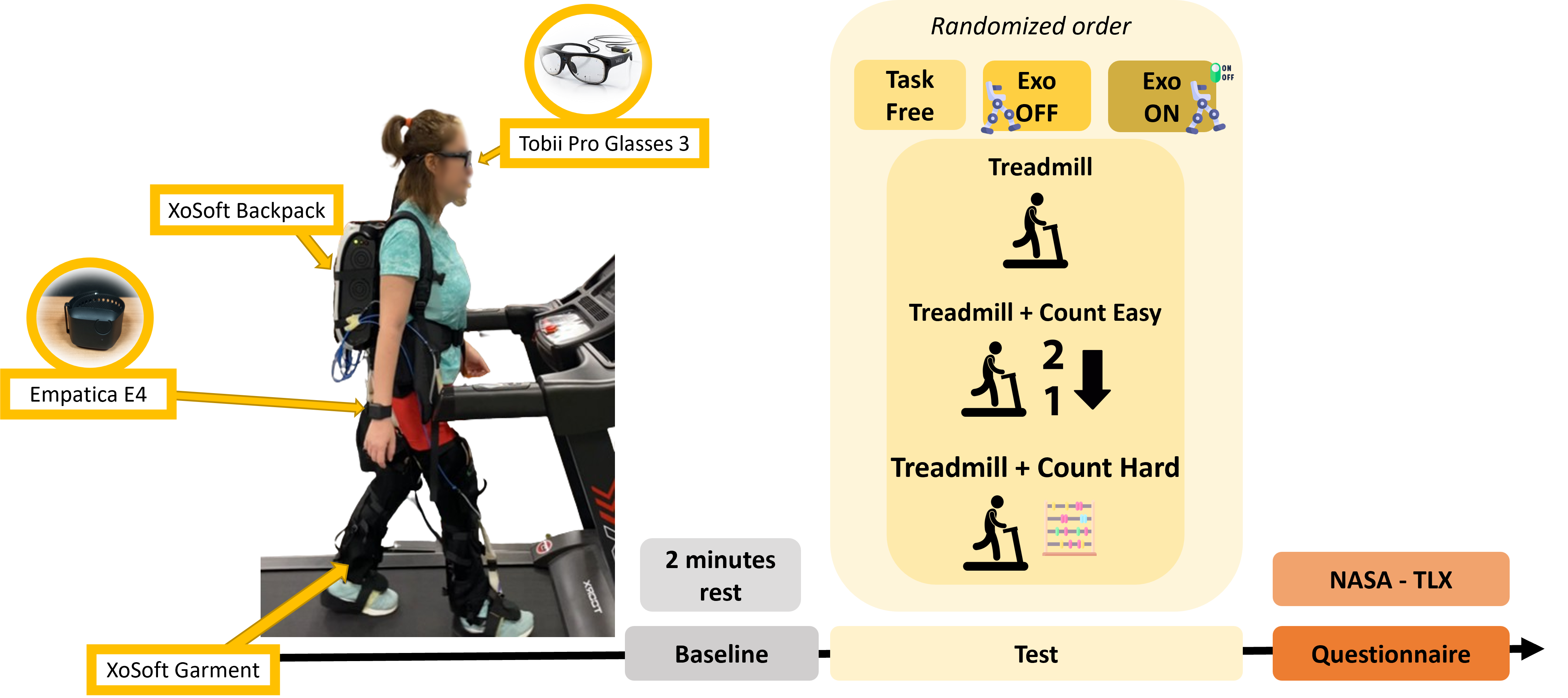}}
    \caption{Experimental setup and procedure. The subject picture shows the used sensors and XoSoft's garment and backpack. The insoles, made of silicone with integrated force-sensitive resistors, are placed inside the shoes. 
    The second part of the figure shows the experimental procedure, displaying all the combinations grouped by exoskeleton conditions. The tasks are randomized and after each one the subject had to answer to the NASA-TLX questionnaire.}
    \label{fig:setup_senors}
\end{figure*}

\subsection{XoSoft Gamma}
XoSoft Gamma, shown in Fig.~\ref{fig:setup_senors}, is a quasi-passive exosuit designed for lower limb assistance. It was designed for clinical populations \cite{dinatali2020} and was adapted to maintain the muscular tone of astronauts \cite{DiNatali2021}.
It is composed of several components: two leggings with straps to secure the garment and attach the actuators, a 4.4\ kg backpack that houses the electronics and pneumatic system, two insoles embedded with force-sensitive resistors serve as sensors for the finite-state machine that performs gait segmentation.

The actuation strategy relies on storing and releasing elastic energy via selective elongation of an elastic band, thanks to a series actuator comprising a pneumatic clutch and a spring \cite{dinatali2020}. When it is desired to store or release energy from the spring with increased stiffness, the clutch is blocked by creating a lower pressure in it than the atmospheric pressure and thus causing the two comb-like components to interlock. On the other hand, when one does not want to take advantage of the spring, the clutch, which has much lower stiffness, is left free to slide.

The system is modular and can assist up to six movements. In the configuration used, it provided symmetric assistance for hip flexion and extension, as well as ankle dorsiflexion.

When the system is inactive, the garment and the backpack are worn by the user without providing assistance and the system is turned off; instead, in the active condition, the assistance is provided. The vacuum generator causes a noise under 50\ dB that is still noticeable and becomes periodic during walking, so the users are aware of the testing condition.

\subsection{Sensors}
The Tobii Pro Glasses 3 is a wearable binocular eye tracker that allows tracking of both eyes, equipped with two chambers for each eye and 16 infrared illuminators, providing a robust eye-tracking signal to environmental artifacts. The system has a built-in camera that captures the experimental scene with a wide 106\ $^{\circ}$ field of view in full HD, including audio recorded from the ambient microphone. An integrated gyroscope and magnetometer also allow for monitoring the subject's movements for cohort detection of eye movements. This eye-tracking system can sample data at 100\ Hz. \\
Empatica E4 is a wristband technology that can be worn on the left hand with photoplethysmography (PPT), electrodermal activity (EDA), accelerometer, and skin temperature sensors. The software automatically generates the following signals: blood volume pulse (BVP) at 64\ Hz, interbeat interval (IBI), heart rate (HR) at 1\ Hz, electrodermal activity at 4\ Hz, raw XYZ acceleration at 32\ Hz, and skin temperature at 4\ Hz. In this study, only the heart rate signal was preliminarily analyzed.

\subsection{Experimental Protocol}
The experiments are conducted on 18 healthy subjects from 24 to 46 years of age (M = 29.7, SD = 5.6), distributed in 10 males and 8 females. All 18 participants completed every test in a randomized order, with each experimental session lasting approximately one hour. All participants gave their written informed consent before the study's commencement, in alignment with the Declaration of Helsinki guidelines. The full experimental procedure adhered to the IIT protocol (001/2019), which received approval from the Ethics Committee of the Liguria Region in Genoa.
Given the exploratory nature of the study and the fact that we were inducing cognitive stress during a walking task, we chose to start with healthy participants, who also offer more homogeneous responses compared to the variability typically found in a population with gait impairments.

\subsection{Experimental Procedure}
Before beginning the experimental protocol, each participant is asked to stand and relax for two minutes to collect baseline physiological signals.

The experiment consists of treadmill walking for two minutes, with the speed standardized at 3.5\ km/h to eliminate speed variation factors, across three different conditions:
\begin{itemize}
\item simple walking;
\item walking while counting down aloud (task \textit{Easy});
\item walking while subtracting 3 aloud (task \textit{Hard}).
\end{itemize}
Because the subjects' native languages are multiple and none of the subjects is a native English speaker, in the two conditions involving a mathematical task, subjects were asked to count in English starting from a three-digit number provided at the beginning of each trial, to avoid any experimental bias and variability among them. The rhythm of the double task was self-paced.
The secondary task was introduced to induce cognitive load in the subject and take mental resources from the primary one, that is walking.

Each of these conditions is conducted in three different modes:
\begin{itemize}
\item \textit{No Exo}: participants complete the conditions walking on the treadmill, without the exosuit;
\item \textit{Exo OFF}: participants perform the three conditions while wearing XoSoft Gamma inactive on the treadmill;
\item \textit{Exo ON}: participants carry out the three conditions while wearing XoSoft Gamma active, which provides assistance to the hip and ankle.
\end{itemize}
The order of conditions is randomized for each participant. Following each condition, participants complete the NASA-TLX, a subjective questionnaire, to measure perceived mental workload during the task.

\section{Metrics}
For the assessment of mental workload, we used both subjective metrics (such as questionnaires) and physiological metrics. In particular, NASA-TLX, pupillometry, and Heart Rate.\\
\subsection{NASA-TLX}
The NASA-TLX is a subjective workload questionary that allows users to assess the subjective workload of operators working with various human-machine interface systems. Developed initially as a paper-and-pencil questionnaire by Sandra Hart of NASA Ames Research Center (ARC) in the 1980s, the NASA TLX has become the gold standard for measuring subjective workload in various applications. Using a multidimensional assessment procedure, the NASA TLX derives an overall workload score based on a weighted average of six subscales: Mental Demand, Physical Demand, Temporal Demand, Performance, Effort, and Frustration.
In this study, we used the raw NASA TLX, so the subscales are not weighted by their perceived importance.

\subsection{Pupillometry}
Pupillometry data was analyzed using Tobii Glasses 3 Eye Tracking, allowing for continuous assessment of pupil dilation in both eyes over time. We performed three-step preprocessing to remove any out-of-band noise, sensory interference, and blink artifacts: first, a fifth-order Butterworth low-pass filter with a cutoff frequency of 4\ Hz was applied \cite{Smallwood}; second, linear interpolation was used to address missing data points \cite{article}; and finally, an amplitude threshold of $ \pm3\sigma$ was implemented to exclude portions of the signal identified as blink artifacts.
Given the high variability in pupil diameter across individuals, an effort was made to standardize the signal by using the percentage change in pupil size (PCPS), calculated according to the following formula \cite{inproceedings}:
\begin{equation} \label{eq_PCPS}
PCPS = \frac{CMPD - BMPD}{BMPD} \cdot 100\%
\end{equation}
where CMPD represents the current measurement of pupil diameter, and BMPD represents the baseline pupil diameter measurement. Based on the PCPS values, the average percentage change in pupil size (APCPS) was then calculated \cite{Aygun}. \\


\subsection{Heart Rate}
All participants' HR was measured with Empatica E4 for various tasks and equipment worn to test general autonomic reactivity using a standard heart rate variability (HRV) geometric method. Geometric methods evaluate the shapes and distribution of RR intervals over the analyzed period. The RR interval represents the length of a ventricular cardiac cycle, measured between two successive R waves, and indicates ventricular rate. For this purpose, a variation curve (histogram of the distribution of intervals) is built, and the main characteristics are determined. Specifically, the Baevsky stress / straining index (SI)~\cite{Chernikova_2017} is calculated according to the formula (\ref{eq_SI}) and expressed as s\textsuperscript{-2}~\cite{ali_2021}.

\begin{equation} \label{eq_SI}
SI = \frac{AMo}{2 \cdot Mo \cdot MxDMn}
\end{equation}

where AMo is the so-called mode amplitude presented in percent, Mo is the mode (the most frequent RR interval) expressed in seconds, and MxDMn is the range of variation reflecting the degree of variability of RR intervals. AMo is obtained as the height of the normalized histogram of RR intervals using a 50\ ms bin width, and MxDMn is the difference between the longest (Mx) and shortest (Mn) RR interval values expressed in seconds.

The index of regulation strain (SI) characterizes the activity of sympathetic or central regulation. Activation of the central loop or sympathetic regulation during mental or physical stresses is manifested by stabilization of rhythm, decrease in interval duration, and increase in the number of intervals of similar duration (AMo growth), therefore, the shape of the histogram can change. In stressful situations and cases of pathology, the diagram will have a narrow base and a sharp (excessive) peak. The asymmetrical diagram may be associated with a transient process and reduced stability. The multimodal diagram indicates a non-sinus rhythm (extrasystole, ciliary arrhythmia). Digitally, it can be represented by the histogram's height-to-width ratio. 

Below, to facilitate a comparison of the study results, instead of using SI, $SI_{dB} = 10 \cdot log(SI)$ is used. Typical SI\textsubscript{dB} fluctuates within 19-22\ dB. It is sensitive to increased sympathetic tone. Mild physical or emotional stress increases SI by 1.5-3\ dB. Severe stress increases SI\textsubscript{dB} by 7-10\ dB. In resting patients with permanently tense regulatory systems, SI\textsubscript{dB} is 26-28\ dB. In resting patients afflicted with attacks of angina pectoris and heart attack, SI\textsubscript{dB} reaches 30-32\ dB~\cite{Chernikova_2017}.

\subsection{Statistical analysis}
The results are statistically analyzed with the Wilcoxon signed rank pairwise test, to assess if there is a significant difference between the testing conditions. 
Since the results present 18 comparisons, the Bonferroni correction was applied. It compensates for this increase by testing each individual hypothesis at a significance level of $\alpha/m$, where $\alpha$ is the overall desired significance level and $m$ is the number of hypotheses. Considering the overall desired level for significance at 5\% and the overall desired level for very significance at 1\%, the corrected significance level becomes 0.28\% and 0.056\%, respectively.


\section{Results}
This section describes the results obtained while the subject is without XoSoft (\textit{No Exo}), wearing XoSoft off (\textit{Exo OFF}) or XoSoft on (\textit{Exo ON}) during a simple 3.5\ km/h walk, with the addition of a countdown aloud (\textit{Easy}) while counting backward in 3-value steps (\textit{Hard}). 
The metrics compared are the NASA-TLX, the average percentage change in pupil size (APCPS), and the Baevsky stress index (SI). 

\subsection{NASA-TLX results}
\begin{figure}
      \centering
	   \begin{subfigure}{0.49\linewidth}
        \caption{No exo}
		\includegraphics[width=\linewidth]{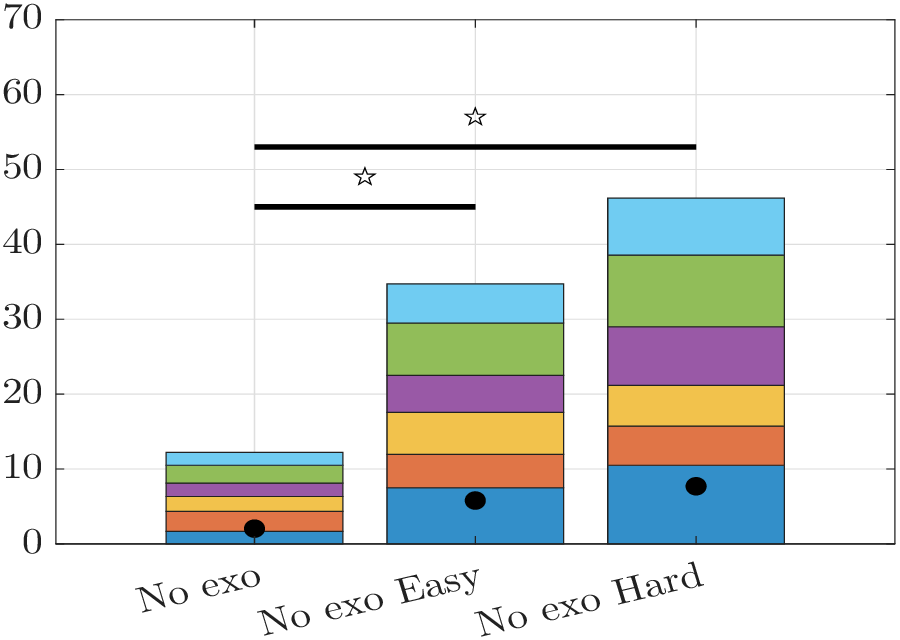}		
		\label{fig:TLX_no_exo}
	   \end{subfigure}
	   \begin{subfigure}{0.49\linewidth}
        \caption{No task}
		\includegraphics[width=\linewidth]{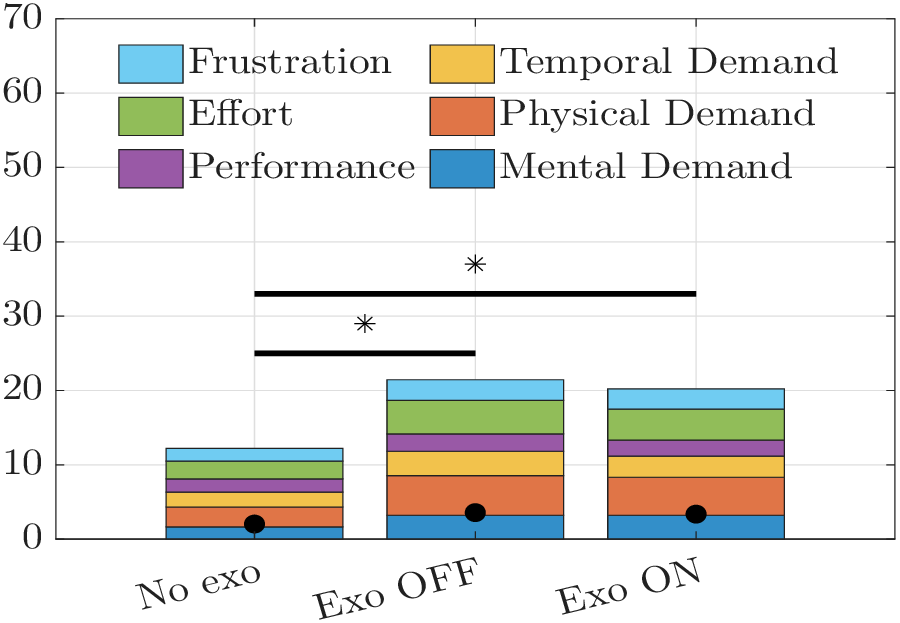}		
		\label{fig:TLX_no_task}
	    \end{subfigure}

	     \begin{subfigure}{0.49\linewidth}
		 \caption{Exo OFF}
		 \includegraphics[width=\linewidth]{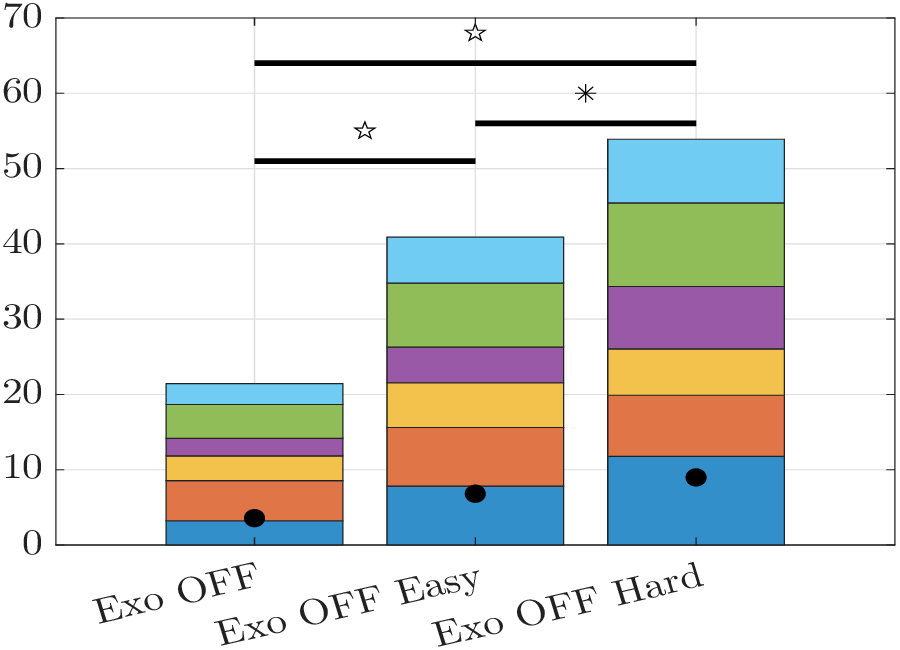}
		 \label{fig:TLX_exo_off}
	      \end{subfigure}
	       \begin{subfigure}{0.49\linewidth}
		  \caption{Task Easy}
		  \includegraphics[width=\linewidth]{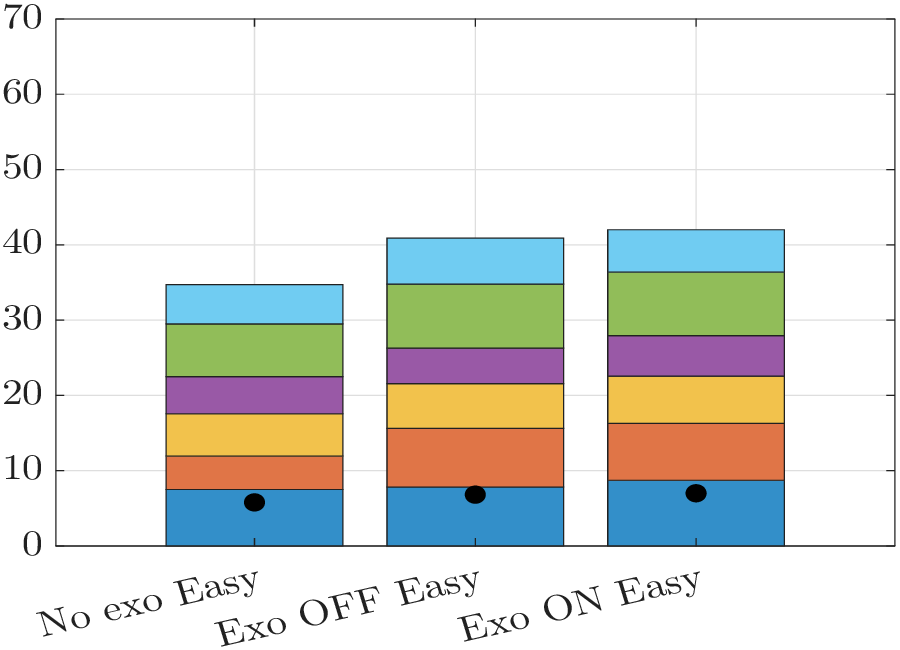}
		  \label{fig:TLX_task_easy}
	       \end{subfigure}
           
	     \begin{subfigure}{0.49\linewidth}
		 \caption{Exo ON}
		 \includegraphics[width=\linewidth]{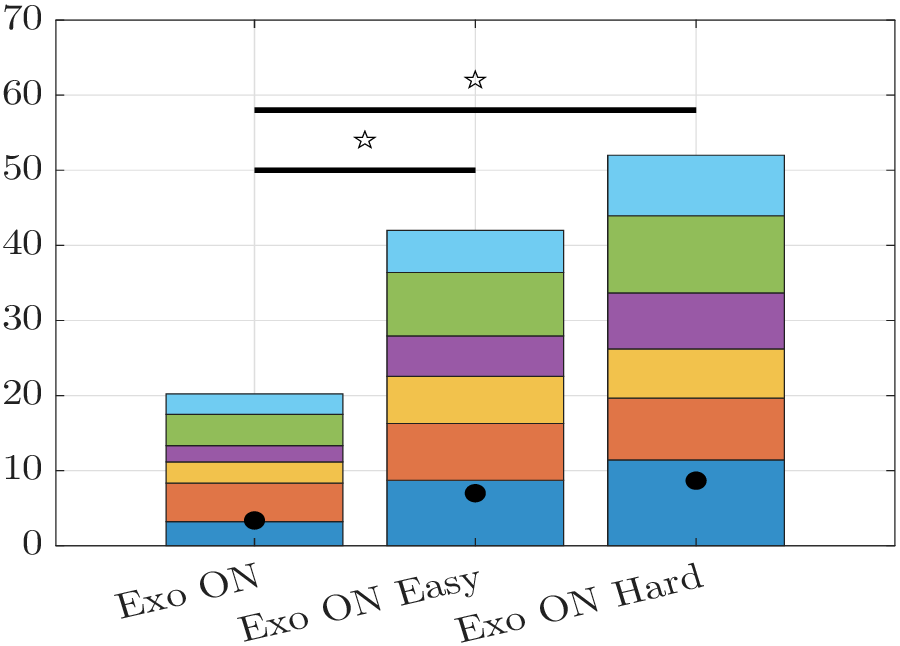}
		 \label{fig:TLX_exo_on}
	      \end{subfigure}
	       \begin{subfigure}{0.49\linewidth}
		  \caption{Task Hard}
		  \includegraphics[width=\linewidth]{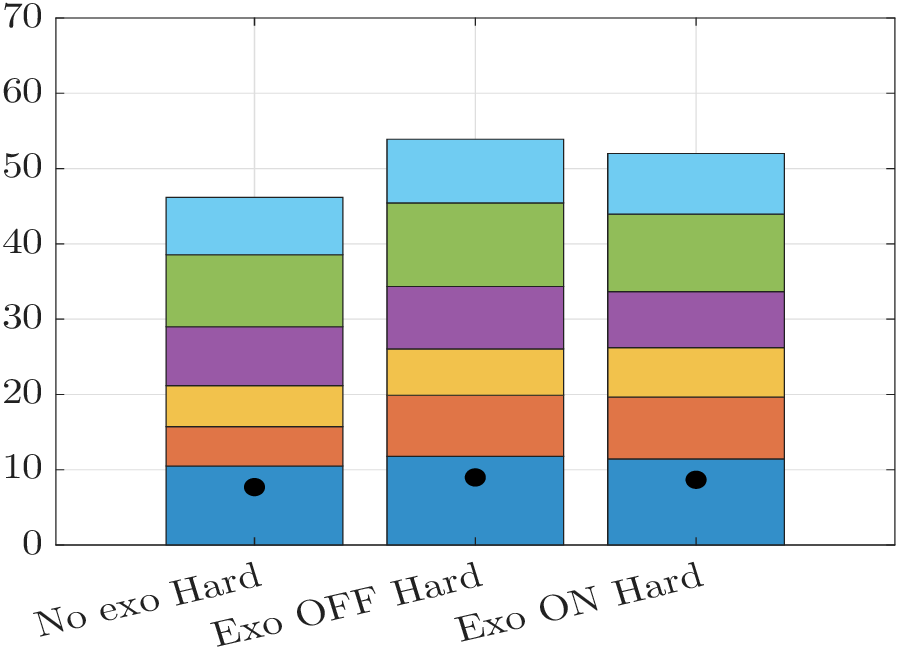}
		  \label{fig:TLX_task_hard}
	       \end{subfigure}
    \caption{Summarized values for all 18 subjects, stacked per activity and divided by color for each field of the NASA-TLX. The first column groups the experiments by exoskeleton condition, and the second column by task. The black dot is the overall TLX value, given by the average of the single fields. The significance refers to the difference between the overall NASA-TLX value distributions. The asterisk reports the significance ($p<0.28\%$) and the five-pointed star the very significance ($p<0.056\%$).}
    \label{fig:NASA_TLX}
\end{figure}

The results of the NASA-TLX questionnaires are reported in Fig.~\ref{fig:NASA_TLX}, which shows the average values of the single fields, the overall NASA-TLX value and the significance of the paired Wilcoxon test.

A first consideration regards the cumulative values of the NASA-TLX questionnaire Fig.~\ref{fig:NASA_TLX}: they result increasingly high according to how challenging the dual task is, with the lower values being consistently those of the simple walking task, and increasing progressively with the task \textit{Easy}, that is the countdown, and the task \textit{Hard}, that is the subtractions by 3. Both for the free walking condition and the count hard condition, the load index of \textit{Exo OFF} is higher than \textit{Exo ON}, while this is not the case for the count easy condition.

Higher results for the \textit{Exo OFF} condition suggest that expectation about the assistance provided by the exoskeleton has an impact on how workload is perceived by the users. On the other hand, the noise caused by the pneumatic system seems not to have been a stress factor during the experiment. The questionnaire components with the highest value for the simple walking with the tasks appear to be \textit{Effort} and \textit{Physical Demand}. Both of them could be related to the weight of the backpack and to the burden added while wearing the device.

The paired Wilcoxon test was used to assess the significance of the differences seen in the paired conditions. 
For the \textit{No Exo} condition, there is high significance between the simple walking and the \textit{Easy} ($p = 0.020\%$) and \textit{Hard} ($p = 0.019\%$) tasks, while the is no significant difference between the two mathematical tasks. The \textit{Exo OFF} condition results show a very significant difference between  simple walking and the \textit{Easy} ($p = 0.020\%$), simple walking and \textit{Hard} ($p = 0.020\%$), but also show a significant difference between the double-task conditions ($p = 0.11\%$). The results of the \textit{Exo ON} condition recall those of the \textit{No Exo} one (respectively $p = 0.020\%$, $p = 0.023\%$).
The statistical test results are different when comparing the same task under different exoskeleton conditions. As shown in the second column of Fig.~\ref{fig:NASA_TLX} the perceived workload is influenced by the presence of the exosuit only when the task is simple walking, with a significant difference between the \textit{No Exo} and \textit{Exo OFF} ($p = 0.12\%$) and the \textit{No Exo} and \textit{Exo ON} ($p = 0.21\%$). Instead, when the task becomes more complex, the perceived workload seems to be primarily associated with it, while the presence or lack of XoSoft does not significantly affect the results.


\subsection{Average Percentage Change in Pupil Size (APCPS)}
\begin{figure}
      \centering
	   \begin{subfigure}{0.49\linewidth}
        \caption{No exo}
		\includegraphics[width=\linewidth]{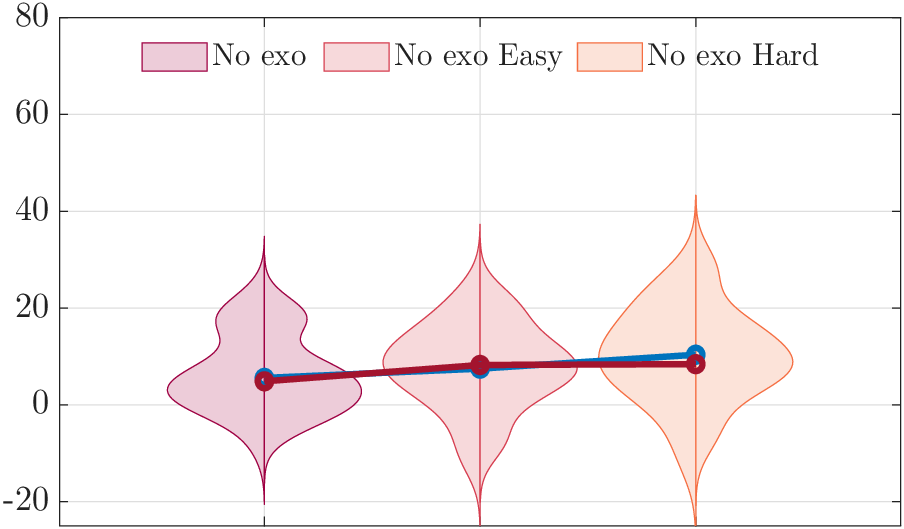}		
		\label{fig:APCPS_no_exo}
	   \end{subfigure}
	   \begin{subfigure}{0.49\linewidth}
        \caption{No task}
		\includegraphics[width=\linewidth]{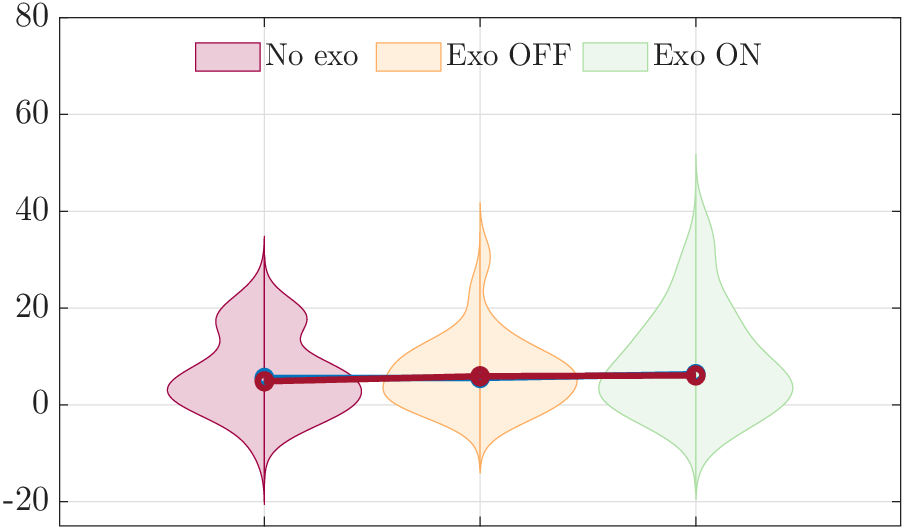}		
		\label{fig:APCPS_no_task}
	    \end{subfigure}

	     \begin{subfigure}{0.49\linewidth}
		 \caption{Exo OFF}
		 \includegraphics[width=\linewidth]{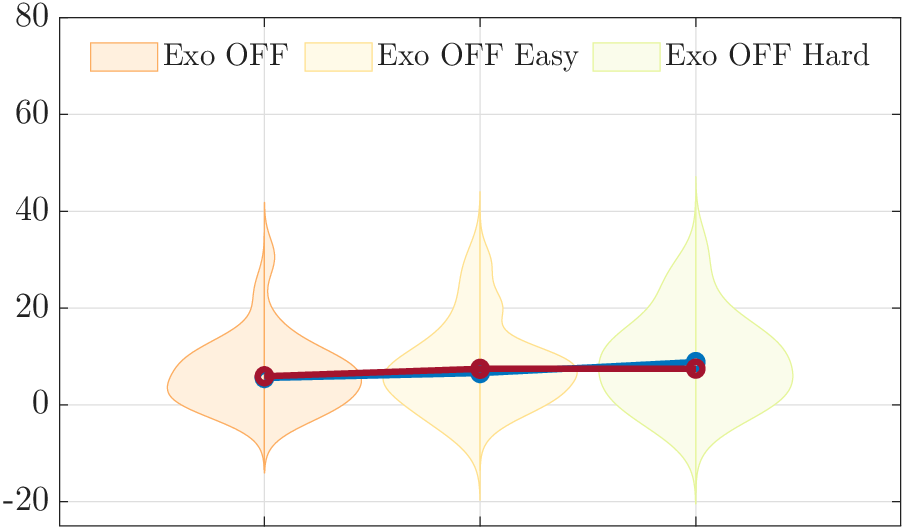}
		 \label{fig:APCPS_exo_off}
	      \end{subfigure}
	       \begin{subfigure}{0.49\linewidth}
		  \caption{Task Easy}
		  \includegraphics[width=\linewidth]{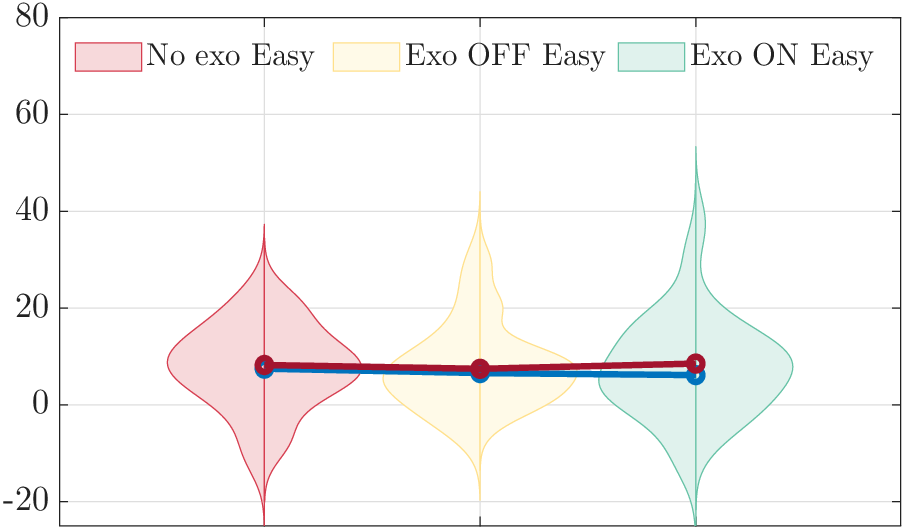}
		  \label{fig:APCPS_task_easy}
	       \end{subfigure}
           
	     \begin{subfigure}{0.49\linewidth}
		 \caption{Exo ON}
		 \includegraphics[width=\linewidth]{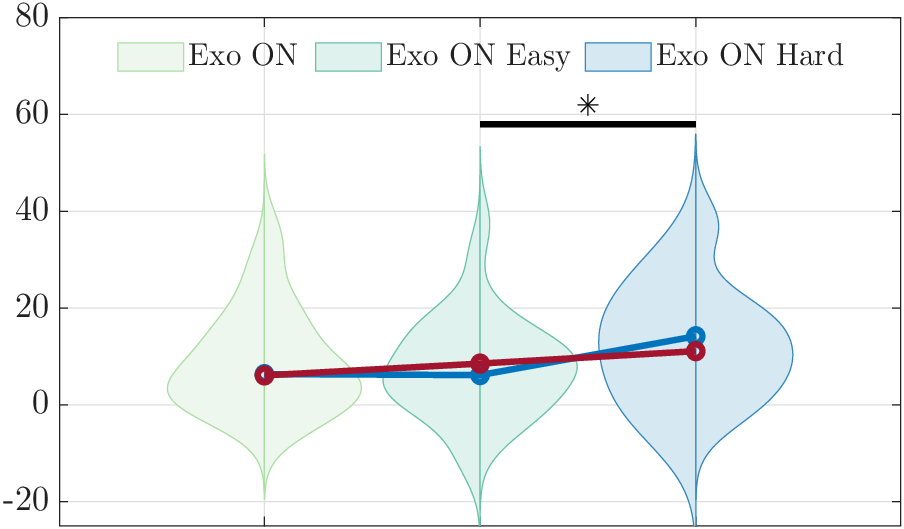}
		 \label{fig:APCPS_exo_on}
	      \end{subfigure}
	       \begin{subfigure}{0.49\linewidth}
		  \caption{Task Hard}
		  \includegraphics[width=\linewidth]{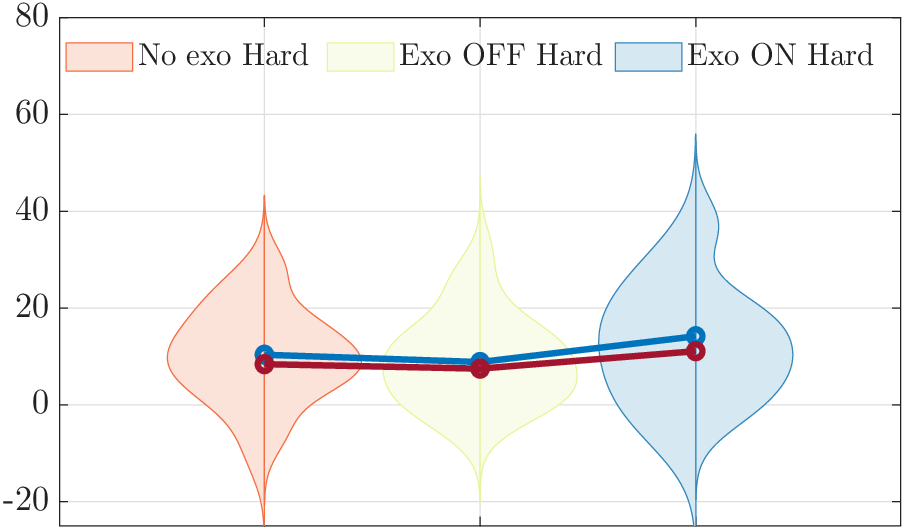}
		  \label{fig:APCPS_task_hard}
	       \end{subfigure}
    \caption{Average Percentage Change in Pupil Size (APCPS) distributions. The first column groups the experiments by exoskeleton condition, and the second column by task. Each combination has a unique color that is maintained. The left half-portion of each violin plot is related to the left eye, while the right half-portion is related to the right eye. Blue color refers to the left eye and red to the right eye. The lines describe the trend of the median value of the data distributions. The asterisk reports the significance ($p<0.28\%$) and the five-pointed star the very significance ($p<0.056\%$).}
    \label{fig:APCPS}
\end{figure}

The analysis of the APCPS shows significant differences across conditions, tasks, and eyes, as how shown in Fig.~\ref{fig:APCPS} and supported by statistical analysis. As the first column shows, the mean values of APCPS appear to increase with the increasing difficulty of the tasks: with lower values observed in the simple walking conditions and higher values in the harder task conditions. This trend is not observed with the exoskeleton configurations, that present different mean values according to the task to perform, similarly to what happened with the NASA-TLX results. Both the left eye value (negative side of the violin) and the right eye value (positive side) are shown paired in the figure; a Wilcoxon test was performed for the right eye, the only significance revealed is between \textit{Exo ON Easy} and \textit{Exo ON Hard} ($p = 0.086\%$) and this result highlights the influence of the use of the exoskeleton and the complexity of the task on pupil size, particularly under the most difficult working conditions.
 For the left eye, with the same statistical analysis, no significant differences are observed. This could be an interesting result showing the different responses between two eyes under different conditions.

\subsection{Baevsky stress index (SI)}

\begin{figure}
      \centering
	   \begin{subfigure}{0.49\linewidth}
        \caption{No exo}
		\includegraphics[width=\linewidth]{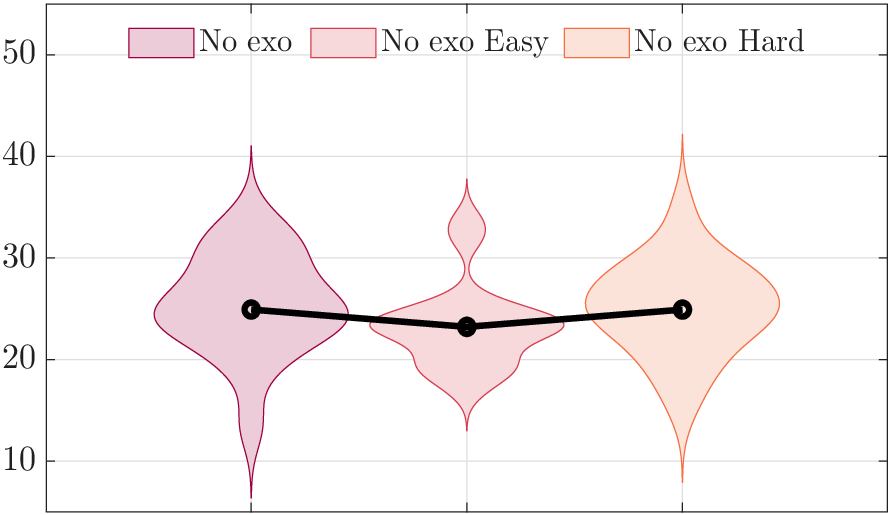}		
		\label{fig:SI_no_exo}
	   \end{subfigure}
	   \begin{subfigure}{0.49\linewidth}
        \caption{No task}
		\includegraphics[width=\linewidth]{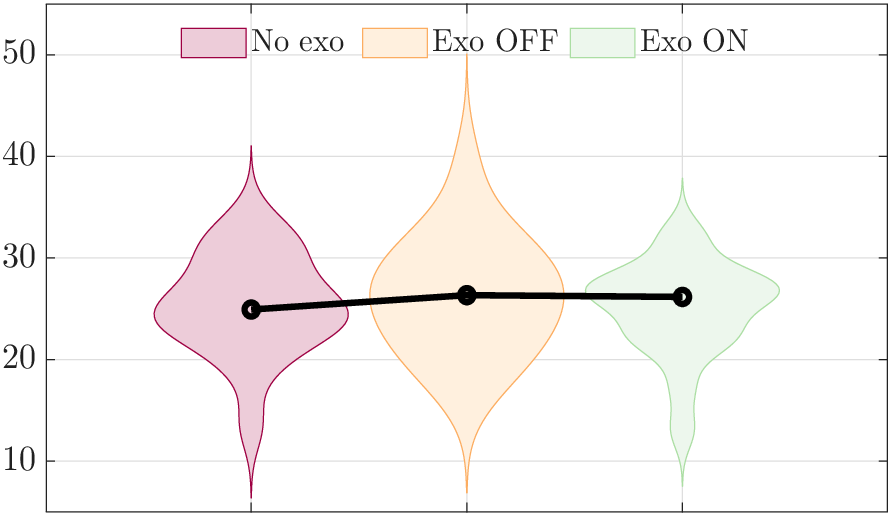}		
		\label{fig:SI_no_task}
	    \end{subfigure}

	     \begin{subfigure}{0.49\linewidth}
		 \caption{Exo OFF}
		 \includegraphics[width=\linewidth]{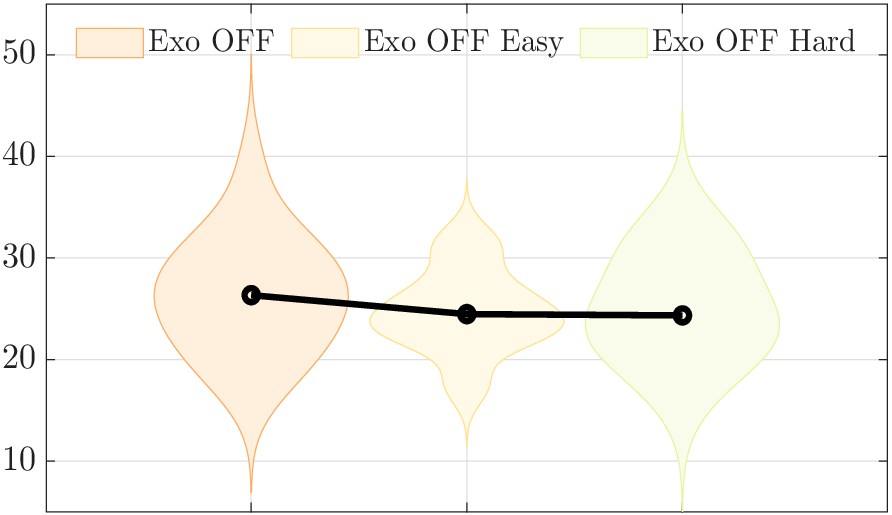}
		 \label{fig:SI_exo_off}
	      \end{subfigure}
	       \begin{subfigure}{0.49\linewidth}
		  \caption{Task Easy}
		  \includegraphics[width=\linewidth]{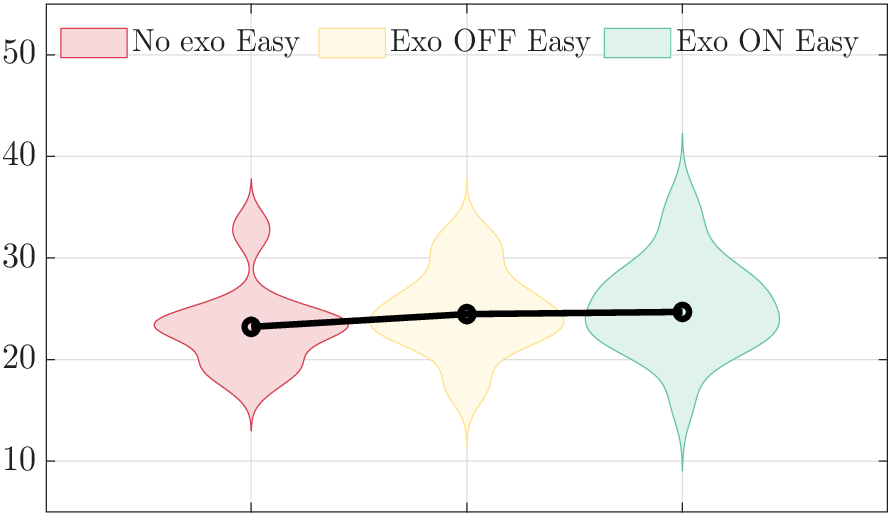}
		  \label{fig:SI_task_easy}
	       \end{subfigure}
           
	     \begin{subfigure}{0.49\linewidth}
		 \caption{Exo ON}
		 \includegraphics[width=\linewidth]{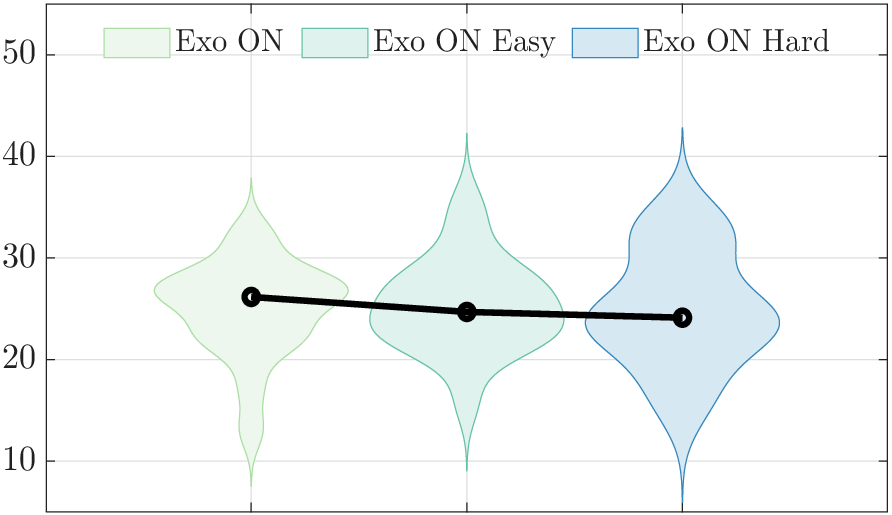}
		 \label{fig:SI_exo_on}
	      \end{subfigure}
	       \begin{subfigure}{0.49\linewidth}
		  \caption{Task Hard}
		  \includegraphics[width=\linewidth]{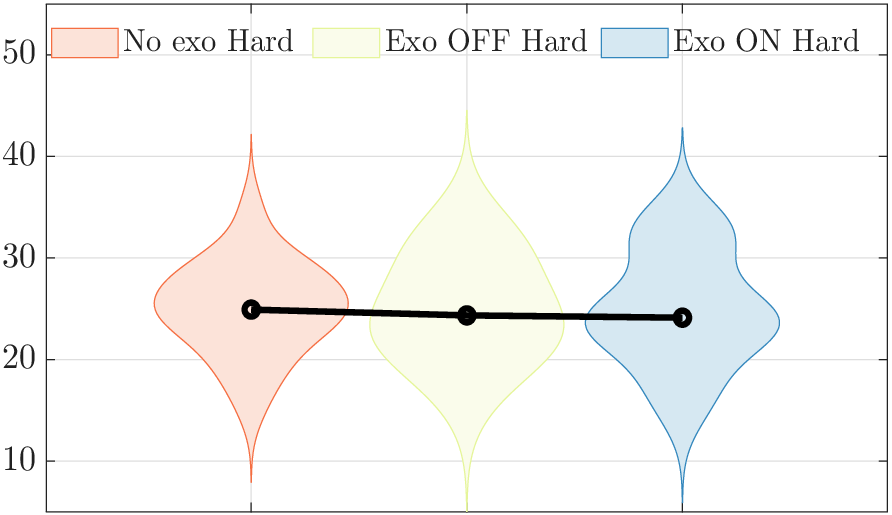}
		  \label{fig:SI_task_hard}
	       \end{subfigure}
    \caption{Baevsky stress index (SI) distributions. The first column groups the experiments by exoskeleton condition, and the second column by task. The lines describe the trend of the median value of the data distributions.}
    \label{fig:SI}
\end{figure}

No statistical significance was derived from the Baevsky stress index data, see Fig.~\ref{fig:SI}. The stress level obtained during the \textit{No Exo Easy} condition is lower than that measured in the absence of the math task or during the most challenging math task. As expected, this suggests that the subjects are more comfortable performing an easy task compared to a difficult task. However, this result is reversed when wearing the exoskeleton. Subjects with both the exoskeleton off and the exoskeleton on have a level of stress that is lower in the hard task than in the easy task, receiving physical and psychological help from the device. The average value of \textit{Exo OFF Hard} and \textit{Exo ON Hard} is lower than that of \textit{No Exo Hard}.

Comparing the conditions for the same required task shows an increasing stress level between the exoskeleton's absence and the \textit{Exo OFF} condition, which could be explained by the dead weight of XoSoft during walking. In the case of \textit{Exo ON} versus \textit{Exo OFF}, on the other hand, the measured stress does not receive a significant increase, showing how the weight, bulk, and noise of XoSoft are offset by the support it provides to the subject. In contrast, evaluating the mean values of \textit{Exo OFF Hard} and \textit{Exo ON Hard}, the latter is slightly lower.

Overall, analyzing the Baevsky stress index provides discordant/complementary indications to those obtained from NASA-TLX and APCPS. It is less sensitive to the MWL induced by different tasks than the other metrics, as the trend of the first column shows. Meanwhile, as the trend of the second column shows, the results related to the exosuit configuration offer a complementary view to the analysis of the previous metrics.

\subsection{Cross-correlation}
Since no statistical significance was derived from the Baevsky stress index data, a more in-depth comparison exclusively between the objective metrics obtained from eye tracking and the subjective values assigned during the completion of the NASA-TLX questionnaire is shown.

\begin{figure*}[h]
    \centering
    \includegraphics[width=\linewidth]{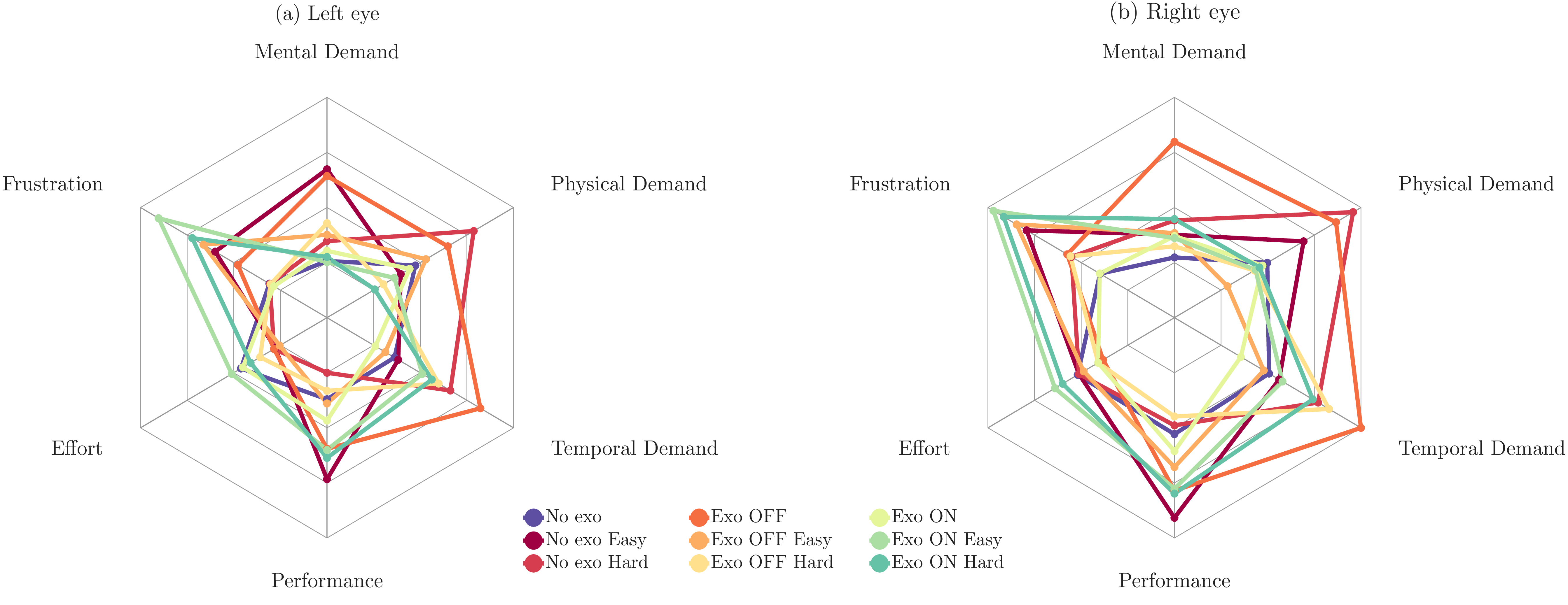}
    \caption{Comparative radar charts illustrating the correlation between APCPS metrics and NASA-TLX workload subscales for different experimental conditions, separated by (a) left eye and (b) right eye. The correlation values are normalized to the maximum correlation value equal to 0.62.}
    \label{fig:correlation}
\end{figure*}

Cross-correlation was performed between the APCPS values and the NASA-TLX results to understand if there is consistency between the two metrics or if they manage to display the same underlying mechanisms.
Fig.~\ref{fig:correlation} shows two radar graphs highlighting the correlation between the APCPS value (for both the left eye and right eye) and the subjective values of the NASA-TLX questionnaire for each field.

Fig.~\ref{fig:correlation}a illustrates the correlation between the left eye and NASA-TLX scores. During more complex tasks (e.g., \textit{No Exo Hard} and \textit{Exo ON Hard}), there are high levels of correlation with Mental Demand, Effort, Frustration, and Physical Demand. This supports the notion that increasing task difficulty significantly elevates cognitive load.
In Fig.~\ref{fig:correlation}b, the correlation between the right eye and the questionnaire responses is presented. Under conditions such as \textit{Exo ON Hard}, there is an increase in the correlation with the subjective dimensions related to stress, including Mental Demand, Effort, and Frustration. 

The correlation patterns, along with the results presented in the previous sections, align with the concept of oversaturation of cognitive resources, where the added complexity introduced by interacting with the device further challenges the subject's cognitive capacity. Particularly, this effect is more pronounced for the right eye, suggesting it is more sensitive to cognitive overload than the left eye \cite{Dalveren, Poynter02112017}. An example is the Frustration indicator in the right eye, which correlates more with the values acquired with \textit{Exo ON} with respect to the same task without the exoskeleton Fig.~\ref{fig:correlation}b.

The results confirm the previous observation that pupillary dynamics in the right eye show a stronger correlation with NASA-TLX responses than those in the left eye. This difference may indicate a functional asymmetry between the eyes, potentially reflecting differences in the brain's cognitive processing or dominance of the right eye, which is often linked to motor control and handling complex information \cite{Asymmetry, ooi2020sensory}.

\section{Discussion}
The results obtained from the APCPS analysis suggest that the use of the exoskeleton and the different levels of task difficulty significantly influence the subject's pupil diameter. They also highlight a different response in the dynamics of the right eye compared to the left eye. This ocular difference could reflect a phenomenon of cognitive and/or sensory overload linked to the simultaneous interaction of challenging tasks and the use of a rehabilitation device.

For the right eye, a significant result is observed under the most complex condition: the combination of task difficulty and exoskeleton activation (e.g., \textit{Exo ON Hard}) seems to lead to such a high activation level that it indicates the cognitive saturation of the subject. This result could be interpreted as the outcome of a “double load”: on the one hand, the cognitive system must manage the complex task, while on the other, the use of the exoskeleton requires additional resources for control. This dual-task scenario explains the increase in the APCPS index, particularly evident in challenging tasks with \textit{Exo ON} compared to conditions without the exoskeleton or with simple tasks \cite{Wel}. No significant differences were found in the left eye. However, the difference between the two eyes is also evident from Fig.~\ref{fig:correlation} where the left eye, compared to the right eye, has pupillary dynamics that are less pronounced. This could suggest a functional asymmetry between the two eyes related to differences in information processing by the brain or a dominant effect of the right eye \cite{Dalveren, Poynter02112017}.

The data suggest that activating the exoskeleton triggers a high level of user activation. This is beneficial for simple tasks, providing motor support and facilitating task execution. However, this same activation level becomes detrimental for complex tasks, leading to cognitive overload. In such cases, the user's mental resources are fully engaged, resulting in increased stress and fatigue and ultimately reducing the overall efficiency of the human-machine system.

This trend is also visible in the NASA-TLX results, that show an increase in the overall NASA-TLX when XoSoft is worn. 
For simple walking, the perceived mental workload becomes significantly higher when wearing the exosuit. For the double tasks, the difference between the exoskeleton conditions is reduced, but it has still higher results for \textit{Exo OFF} and \textit{Exo ON}, possibly indicating that the excess of mental resources is harder to self-assess for the users.

A higher correlation of these configurations with the pupil's diameter in the stress-related values highlights the possibility that they are both expressions of oversaturation. 
Finally, the indications obtained from the heart rate and the resulting stress index may advance how different metrics respond to different stimuli and dynamics, suggesting the possibility that a given level of mental workload may be reflected differently at the ocular or cardiac levels. 

\section{Conclusion}
This work introduced a comparison of different physiological metrics with questionnaires for assessing cognitive load introduced by exoskeletons.

A limitation of this study is that the performance data of the double-task experiments are absent. This choice aligns with the study's primary objective of comparing the gold-standard workload assessment protocol with physiological metrics, where behavioral performance metrics would not directly contribute to this specific validation. Future work will expand on these aspects by integrating performance-psychophysiological relationships.

The results show a complex interaction between task difficulty, exoskeleton activation, and pupillary dynamics, suggesting that the subject might reach a saturated condition under high load. This phenomenon opens interesting perspectives for optimizing exoskeleton design and understanding the dynamics of cognitive and physical overload in complex contexts.

The data suggest that pupil diameter may be an objective cognitive load indicator that correlates with subjective NASA-TLX questionnaires. The results confirm that pupillary dynamics in the right eye show a stronger correlation with NASA-TLX responses than those in the left eye. This ocular asymmetry highlights that the mechanisms of cognitive load are not uniform and may depend on both physiological factors (lateralization) and the nature of the task (mathematical, visual, and motor).
In the future, we plan to expand the number of subjects to test the hypotheses presented, particularly those regarding ocular metrics. At the same time, the reason for the lack of significance for cardiac metrics will have to be verified.

From an exoskeleton design point of view, the results suggest that the current prototype increases the users' perceived mental workload for its primary task, which is walking assistance.
The results on physical demand show the impact of the backpack weight, suggesting that the next prototype should address this issue, along with guaranteeing a comfortable garment for the user.
It will be necessary to further investigate the impact of the device on cognitive load, particularly in the target population, to avoid overloading conditions. To see the actual impact of the assistance provided, regardless of the expectation, the experimental design should ensure that the subject is unaware of the device's state. Another factor that will be addressed in future work is the accustoming of the user to the system and if a more prolonged usage of the system might reduce the cognitive load and the effect of the user's expectations on the device, that impact its evaluation.






\bibliographystyle{ieeetr} 
\bibliography{ref}

\end{document}